\documentclass[english]{revtex4}
\usepackage[T1]{fontenc}
\usepackage[latin9]{inputenc}
\usepackage{xcolor}
\usepackage{pdfcolmk}
\usepackage{amsmath}
\usepackage{amssymb}
\usepackage{esint}
\PassOptionsToPackage{normalem}{ulem}
\usepackage{ulem}

\makeatletter

\providecolor{lyxadded}{rgb}{0.99,0.0078,1}
\providecolor{lyxdeleted}{rgb}{1,0,0}

\DeclareRobustCommand{\lyxsout}[1]{\ifx\\#1\else\sout{#1}\fi}

\@ifundefined{textcolor}{}
{%
 \definecolor{BLACK}{gray}{0}
 \definecolor{WHITE}{gray}{1}
 \definecolor{RED}{rgb}{1,0,0}
 \definecolor{GREEN}{rgb}{0,1,0}
 \definecolor{BLUE}{rgb}{0,0,1}
 \definecolor{CYAN}{cmyk}{1,0,0,0}
 \definecolor{MAGENTA}{cmyk}{0,1,0,0}
 \definecolor{YELLOW}{cmyk}{0,0,1,0}
}

\usepackage{graphics}\usepackage{epstopdf}\usepackage{epsfig}\@ifundefined{definecolor}{\usepackage{color}}{}
\usepackage{amsfonts}\usepackage{bm}\usepackage{amscd}

\usepackage{babel}

\usepackage{babel}

\makeatother

\usepackage{babel}
\begin{document}
\title{Non-adiabatic perturbation theory of the exact factorisation }
\author{Matisse Wei-Yuan Tu}
\affiliation{Fritz Haber Center for Molecular Dynamics, Institute of Chemistry,
The Hebrew University of Jerusalem, Jerusalem 91904 Israel}
\author{E.K.U. Gross}
\affiliation{Fritz Haber Center for Molecular Dynamics, Institute of Chemistry,
The Hebrew University of Jerusalem, Jerusalem 91904 Israel}
\begin{abstract}
We present a novel nonadiabatic perturbation theory (NAPT) for correlated
systems of electrons and nuclei beyond the Born-Oppenheimer (BO) approximation.
The essence of the method is to exploit the smallness of the electronic-to-nuclear
mass ratio by treating the electron-nuclear correlation terms in the
electronic equation of motion of the exact factorisation (EF) framework
as perturbation. We prove that any finite-order truncation of the
NAPT preserves the normalisation of the conditional electronic factor
as well as the gauge covariance of the resulting perturbative equations
of motion. We illustrate the usefulness of NAPT by obtaining nonadiabatic
corrections to the BO Berry phase in Jahn--Teller systems with a
conical intersection. It well captures the departure of the exact
Berry phase from being topological via the lowest-order NAPT. By removing
the conical intersection with a constant gap, it further yields the
correct scaling of the Berry phase toward zero. 
\end{abstract}
\maketitle

\section{Introduction}

The correlated electron-nuclear problem is characterised by a nature-given
small dimensionless parameter, the electronic-over-nuclear mass ratio,
\begin{equation}
\mu=\frac{m_{e}}{M},\label{mass-ratio-1}
\end{equation}
in which $m_{e}$ is the electron mass and $M$ is a nuclear reference
mass that can be chosen to be the proton mass $m_{p}$ or the average
nuclear mass of the particular system at hand. $\mu$ is upper bounded
by a small number $\mu\le m_{e}/m_{p}\approx5.4\times10^{-4}$ since
$m_{p}$ is the smallest possible value one can assign to $M$. This
naturally small parameter has been extensively exploited to devise
various approximation schemes for tackling the correlated electron--nuclear
dynamics \cite{Born27457,Hagedorn86571,Nafie925687,Scherrer135305,Scherrer15074106,Eich16054110,Axel163316,Scherrer17031035,Panati02250405,Panati07297}.

The diversity of approximation schemes based on this small parameter
reflects the variety of phenomena that motivate them. The specific
question under consideration not only guides the choice of processes
or physical quantities around which the perturbation expansion is
developed, but also determines---on physical grounds---the appropriate
definition of the unperturbed state \cite{Born27457,Hagedorn86571,Nafie925687,Scherrer135305,Scherrer15074106,Eich16054110,Axel163316,Scherrer17031035}.
Furthermore, there exists a distinct class of mathematically oriented
studies that approach the small-$\mu$ limit without explicitly anchoring
the perturbation framework to any particular physical processes or
quantities. Instead, these works focus on analysing how the full solution
to the Schr\"{o}dinger equation depends on $\mu$ and on establishing
rigorous error bounds for various approximation schemes in terms of
their $\mu$-scaling behaviour \cite{Panati02250405,Panati07297},
in the spirit of time-adiabatic analysis \cite{Kato50435}. 

Historically, the seminal work of Born and Oppenheimer \cite{Born27457}
established a systematic approach for exploiting the smallness of
the electron--nuclear mass ratio. By choosing the small vibrational
amplitude of the nuclei around their equilibrium configuration as
the perturbative quantity, they effectively defined an unperturbed
state in which the nuclei are treated as clamped classical points
while the electronic degrees of freedom remain fully quantum. This
prescription underpins the widely used Born--Oppenheimer (BO) approximation,
which provides the foundation for electronic structure theory. Moreover,
by allowing the clamped nuclei to move according to Newtonian dynamics
on the corresponding BO potential energy surface, it naturally supports
mixed quantum--classical approaches \cite{Tully98407,Kapral19998919,Kapral2015073201}.
Notably, the first-order perturbation in this framework (hereafter
abbreviated as BOPT) yields the celebrated energy separation between
electronic states, of order $\mathcal{O}\left(\mu^{0}\right)$, and
nuclear vibrational energies, of order $\mathcal{O}\left(\mu^{1/2}\right)$
\cite{Born27457}. This separation highlights the energetic dominance
of the unperturbed electronic state that underlies the usual electronic
structure studies. 

Importantly, while BOPT fully exploits the smallness of $\mu$, it
is not designed to directly capture nonadiabatic \textit{electronic}
effects that arise when multiple BO potential energy surfaces are
involved. Since $\mu$ enters the full electron--nuclear Schr\"{o}dinger
equation only through the nuclear kinetic energy \cite{Born27457},
naively taking the limit $\mu\rightarrow0$ corresponds to removing
the nuclear kinetic energy. Following this line, perturbation theories
based on the smallness of $\mu$ naturally identify nuclear motion
as the source of nonadiabaticity, and construct corrections by treating
nuclear-motion-related quantities as perturbations. Examples include
approaches that use the nuclear kinetic energy itself \cite{Diestler20134698},
or the nuclear velocity, as in the nuclear velocity perturbation theory
(NVPT) \cite{Nafie925687,Ditler20222448,Scherrer135305}. Thus, while
BOPT provides a sound physical basis for separating electrons and
nuclei by energy scale, a framework that can simultaneously account
for nonadiabatic effects and preserve such a separation is highly
desirable. Achieving this, however, is far from straightforward when
working directly with the full electron--nuclear Schr\"{o}dinger
equation, whose solutions generally depend inseparably on all degrees
of freedom.

\begin{figure}[h] \includegraphics[width=17cm, height=3.5 cm]{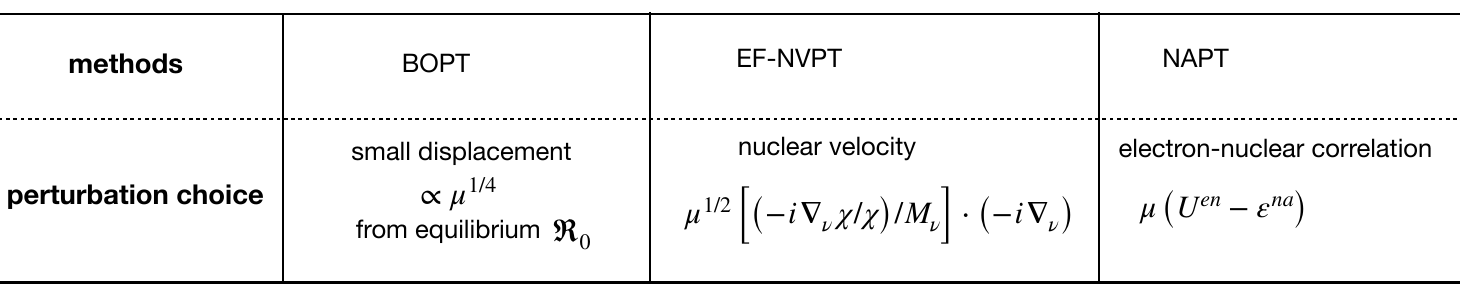} 
\caption{Overview of interrelated perturbation methods exploiting the smallness of the electronic-over-nuclear mass parameter $\mu$.  Here "EF" abbreviates exact factorisation and "PT" denotes perturbation theory, while BOPT, EF-NVPT, and NAPT refer to Born-Oppenheimer, EF-based nuclear-velocity, and electronic-nonadiabatic PT, respectively. } 
\label{regimes} 
\end{figure}

The exact factorisation (EF) provides a natural framework to realise
this separation. While remaining formally exact, EF decomposes the
full electron--nuclear Schrödinger equation into a set of two coupled
equations of motion (EOMs) for the electronic and nuclear subsystems
\cite{Abedi10123002,Suzuki14040501,Min15073001,Li2022113001,Arribas2024233201,Abedi1222A530,Min2014263004,Requist16042108,Requist17062503}.
This decomposition offers a natural starting point for approximating
the two subsystems separately, while fully retaining their mutual
coupling. By combining EF with the existing idea of treating nuclear
velocity as a perturbative quantity \cite{Nafie925687,Scherrer135305},
an EF-based perturbation theory (EF-NVPT) has been formulated \cite{Scherrer15074106}.
This approach has proven computationally efficient in calculating
vibrational circular dichroism \cite{Scherrer15074106} and nuclear-motion-induced
electronic fluxes \cite{Axel163316}. It also offers a rigorous framework
for defining the mass that governs vibrational spectroscopic signatures
in molecules \cite{Scherrer17031035}. Notably, EF-NVPT is rooted
in the perturbative structure naturally revealed by the EF framework:
the small parameter $\mu$ enters explicitly in the nonadiabatic electron--nuclear
correlation term of the electronic EOM. The EF-NVPT does not employ
this entire term as the perturbation; instead, it focuses on using
the nuclear velocity as the organising principle for its expansion.
This naturally suggests an alternative line of development, namely
to construct a perturbation theory that treats the full electron--nuclear
correlation term itself as the perturbation. We refer to this approach
as the electronic--nonadiabatic perturbation theory (NAPT). A brief
comparison of BOPT, EF-NVPT and NAPT is provided in Fig. \ref{regimes}. 

The remainder of this article is organised as follows. Sec. \ref{main-bodytext-ENAPT}
introduces the formulation of NAPT, and Sec. \ref{main-bodytext-ENAPT-postpert}
discusses its structural properties inherited from the EF framework,
in particular the freedom to choose approximations for the nuclear
DOF. These features underlie the application in Ref. \cite{Tu2025043075},
where first-order NAPT combined with classical nuclei was shown to
capture electronic decoherence dynamics along a single nuclear trajectory
\cite{FootnoteENAPTapp1}. Sec. \ref{main-bodytext-ENAPT-exactcond}
proves that finite-order truncations preserve two essential EF properties---the
partial normalisation condition and gauge covariance \cite{Abedi10123002,Suzuki14040501,Min15073001,Li2022113001,Arribas2024233201,Abedi1222A530,Min2014263004,Requist16042108,Requist17062503}---ensuring
consistency with the exact theory. Sec. \ref{examples-BerryPhase}
illustrates the usefulness of NAPT by applying it to calculations
of the Berry phase in Jahn--Teller systems featuring the presence
of conical intersections (see Fig. \ref{BerryCorr} for a quick overview
of the results). Sec. \ref{conclu} is for discussion and outlook.

\section{An electronic-nonadiabatic perturbation theory }

\label{main-bodytext}

The Hamiltonian of a general system of interacting electrons and nuclei
is given by 
\begin{equation}
H=T_{n}+H^{BO}\left(\boldsymbol{R}\right),\label{HtnBO}
\end{equation}
where $T_{n}$ is the nuclear kinetic energy operator and $H^{BO}\left(\boldsymbol{R}\right)$
is the Born-Oppenheimer (BO) Hamiltonian, 
\begin{equation}
H^{BO}\left(\boldsymbol{R}\right)=T_{e}+V_{ee}+V_{en}\left(\boldsymbol{R}\right)+V_{nn}\left(\boldsymbol{R}\right),\label{HBOe}
\end{equation}
which consists of the electronic kinetic energy $T_{e}$, the electronic
Coulomb interaction $V_{ee}$, the Coulomb interactions between electrons
and nuclei $V_{en}\left(\boldsymbol{R}\right)$ and the inter-nuclear
Coulomb interaction $V_{nn}\left(\boldsymbol{R}\right)$. Here $H^{BO}\left(\boldsymbol{R}\right)$
describes the pure electronic structure problem, namely, 
\begin{equation}
H^{BO}\left(\boldsymbol{R}\right)\left\vert \varphi_{k}\left(\boldsymbol{R}\right)\right\rangle =\varepsilon_{k}\left(\boldsymbol{R}\right)\left\vert \varphi_{k}\left(\boldsymbol{R}\right)\right\rangle ,\label{twlvl-Hilbert}
\end{equation}
where the BO eigenenergies $\varepsilon_{k}\left(\boldsymbol{R}\right)$
and eigenfunctions $\left\vert \varphi_{k}\left(\boldsymbol{R}\right)\right\rangle $
are obtained at each fixed nuclear configuration $\boldsymbol{R}=\left\{ \boldsymbol{R}_{\nu}\right\} $
with $\nu$ enumerating the nuclei in the system. By applying the
atomic (Hartree) unit (which will be used from now on), the small
dimensionless parameter $\mu$ of Eq. (\ref{mass-ratio-1}) naturally
appears, namely, $T_{n}=\mu\sum_{\nu}\left(-i\boldsymbol{\nabla}_{\nu}\right)^{2}/\left(2M_{\nu}\right)$
with $M_{\nu}$, the mass of the $\nu$th nucleus, now measured in
terms of the reference mass $M$. Here $\boldsymbol{\nabla}_{\nu}=\partial/\partial\boldsymbol{R}_{\nu}$.
With the appearance of $\mu$ made explicit, a number of different
perturbation analysis starting from the full time-dependent Schr\"{o}dinger
equation (TDSE) $i\left\vert \dot{\Psi}\left(t\right)\right\rangle =H\left\vert \Psi\left(t\right)\right\rangle $
have been developed in various works \cite{Born27457,Hagedorn86571,Scherrer15074106,Nafie925687,Panati02250405,Panati07297}. 

\subsection{The EF and the essential properties of the NAPT}

\label{main-bodytext-ENAPT}

As we set out to develop a perturbation theory where approximations
on top of the perturbation can be done independently for electrons
and nuclei, it is natural to pursue the EF framework. It inherently
factorises the fully correlated wave function, $\Psi\left(\boldsymbol{r},\boldsymbol{R},t\right)$,
of interacting electrons and nuclei, into a product, 

\begin{equation}
\Psi\left(\boldsymbol{r},\boldsymbol{R},t\right)=\chi\left(\boldsymbol{R},t\right)\phi\left(\boldsymbol{r}\left\vert t,\boldsymbol{R}\right.\right),\label{exfc-gen1}
\end{equation}
of a nuclear factor $\chi\left(\boldsymbol{R},t\right)$, also called
the nuclear wave function with an electronic factor $\phi\left(\boldsymbol{r}\left\vert t,\boldsymbol{R}\right.\right)$,which
parametrically depends on $\left(t,\boldsymbol{R}\right)$. Here $\boldsymbol{r}=\left\{ \boldsymbol{r}_{j}\right\} $
denotes the set of electronic coordinates. The factors $\chi\left(\boldsymbol{R},t\right)$
and $\phi\left(\boldsymbol{r}\left\vert t,\boldsymbol{R}\right.\right)$
follow separate, but coupled EOMs \cite{Abedi10123002,Suzuki14040501,Min15073001,Li2022113001,Arribas2024233201,Abedi1222A530,Min2014263004,Requist16042108,Requist17062503}.
The nuclear wave function follows an EOM that looks like an ordinary
TDSE, namely, 

\begin{equation}
i\partial_{t}\chi\left(\boldsymbol{R},t\right)=\left\{ \mu\sum_{\nu}\frac{\left[\left(-i\boldsymbol{\nabla}_{\nu}\right)+\boldsymbol{A}_{\nu}\left[\phi\right]\left(\boldsymbol{R},t\right)\right]^{2}}{2M_{\nu}}+\varepsilon\left[\phi\right]\left(\boldsymbol{R},t\right)\right\} \chi\left(\boldsymbol{R},t\right),\label{EXF-EOM-nu}
\end{equation}
in which 
\begin{equation}
\varepsilon\left[\phi\right]\left(\boldsymbol{R},t\right)=\varepsilon^{A}\left[\phi\right]\left(\boldsymbol{R},t\right)+\mu\varepsilon^{NA}\left[\phi\right]\left(\boldsymbol{R},t\right),\label{EXF-pot-scalsum}
\end{equation}
and
\begin{equation}
\boldsymbol{A}_{\nu}\left[\phi\right]\left(\boldsymbol{R},t\right)=\left\langle \phi\left(t,\boldsymbol{R}\right)\left\vert -i\boldsymbol{\nabla}_{\nu}\phi\left(t,\boldsymbol{R}\right)\right.\right\rangle ,\label{EXF-pot-vec}
\end{equation}
play the same role as the electromagnetic scalar and vector potentials
for charged particles. We adopted the Dirac notation for the electronic
factor, namely, $\phi\left(\boldsymbol{r}\left\vert t,\boldsymbol{R}\right.\right)=\left\langle \boldsymbol{r}\left\vert \phi\left(t,\boldsymbol{R}\right)\right.\right\rangle $.
The scalar potential consists of an adiabatic (A) functional
\begin{equation}
\varepsilon^{A}\left[\phi\right]\left(\boldsymbol{R},t\right)=\left\langle \phi\left(t,\boldsymbol{R}\right)\left\vert \left(H^{BO}\left(\boldsymbol{R}\right)-i\frac{\partial}{\partial t}\right)\right\vert \phi\left(t,\boldsymbol{R}\right)\right\rangle ,\label{EXF-pot-BO}
\end{equation}
and a non-adiabatic (NA) functional
\begin{equation}
\varepsilon^{NA}\left[\phi\right]\left(\boldsymbol{R},t\right)=\left\langle \phi\left(t,\boldsymbol{R}\right)\right\vert U_{K}^{en}\left[\phi\right]\left(\boldsymbol{R},t\right)\left\vert \phi\left(t,\boldsymbol{R}\right)\right\rangle ,\label{EXF-pot-scalgeoFull}
\end{equation}
where
\begin{equation}
U_{K}^{en}\left[\phi\right]\left(\boldsymbol{R},t\right)=\sum_{\nu}\frac{\left[\left(-i\boldsymbol{\nabla}_{\nu}\right)-\boldsymbol{A}_{\nu}\left[\phi\right]\left(\boldsymbol{R},t\right)\right]^{2}}{2M_{\nu}}.\label{EXF-enucor-K}
\end{equation}
The BO potential energy surface is obtained by inserting the BO state
$\varphi_{k}$ into the functional $\varepsilon^{A}\left[\varphi_{k}\right]\left(\boldsymbol{R},t\right)=\varepsilon_{k}\left(\boldsymbol{R}\right)$
while $\varepsilon^{NA}\left[\varphi_{k}\right]\left(\boldsymbol{R},t\right)=\sum_{\nu}\left\langle \varphi_{k}\left(\boldsymbol{R}\right)\right\vert \left[\left(-i\boldsymbol{\nabla}_{\nu}\right)-\boldsymbol{A}_{\nu}\left[\varphi_{k}\right]\left(\boldsymbol{R}\right)\right]^{2}\left\vert \varphi_{k}\left(\boldsymbol{R}\right)\right\rangle /\left(2M_{\nu}\right)$
is the well-known diagonal correction to the BO potential energy surface.

The time evolution of the exact electronic factor follows a TDSE-like
equation \begin{subequations}\label{e-EOM-set}
\begin{equation}
i\partial_{t}\left\vert \phi\left(t,\boldsymbol{R}\right)\right\rangle =\left[H^{BO}\left(\boldsymbol{R}\right)-\varepsilon^{A}\left[\phi\right]\left(\boldsymbol{R},t\right)+V^{en}\left[\phi,\chi\right]\left(\boldsymbol{R},t\right)\right]\left\vert \phi\left(t,\boldsymbol{R}\right)\right\rangle ,\label{e-EOM-full}
\end{equation}
where
\begin{equation}
V^{en}\left[\phi,\chi\right]\left(\boldsymbol{R},t\right)=\mu\left(U^{en}\left[\phi\right]\left(\boldsymbol{R},t\right)-\varepsilon^{NA}\left[\phi\right]\left(\boldsymbol{R},t\right)\right).\label{Venmu}
\end{equation}
\end{subequations} Here 
\begin{equation}
U^{en}\left[\phi\right]\left(\boldsymbol{R},t\right)=U_{K}^{en}\left[\phi\right]\left(\boldsymbol{R},t\right)+U_{Q}^{en}\left[\phi,\chi\right]\left(\boldsymbol{R},t\right)\label{Uen-def}
\end{equation}
 is called the electron-nuclear correlation operator with 
\begin{equation}
U_{Q}^{en}\left[\phi,\chi\right]\left(\boldsymbol{R},t\right)=\sum_{\nu}\boldsymbol{\mathfrak{p}}_{\nu}\left[\phi,\chi\right]\left(\boldsymbol{R},t\right)\cdot\frac{\left[-i\boldsymbol{\nabla}_{\nu}-\boldsymbol{A}_{\nu}\left[\phi\right]\left(\boldsymbol{R},t\right)\right]}{M_{\nu}},\label{Uen_Q}
\end{equation}
in which 

\begin{equation}
\boldsymbol{\mathfrak{p}}_{\nu}\left[\phi,\chi\right]\left(\boldsymbol{R},t\right)=\left[-i\frac{\boldsymbol{\nabla}_{\nu}\chi\left(\boldsymbol{R},t\right)}{\chi\left(\boldsymbol{R},t\right)}+\boldsymbol{A}_{\nu}\left[\phi\right]\left(\boldsymbol{R},t\right)\right],\label{pzero-q}
\end{equation}
 is the so-called nuclear momentum function. It is complex in general
and its imaginary part, $\text{Im}\left\{ \boldsymbol{\mathfrak{p}}_{\nu}\right\} =-i\boldsymbol{\nabla}_{\nu}\left\vert \chi\right\vert /\left\vert \chi\right\vert $,
is often called the nuclear quantum momentum \cite{Min15073001,Arribas2024233201}.
Its real part divided by the nuclear mass $M_{\nu}$, on the other
hand, is the nuclear velocity field, $\text{Re}\left\{ \boldsymbol{\mathfrak{p}}_{\nu}\right\} /M_{\nu}=\left(\boldsymbol{\nabla}_{\nu}\text{arg}\left(\chi\right)+\boldsymbol{A}_{\nu}\right)/M_{\nu}=\boldsymbol{J}_{\nu}/\left(\left\vert \chi\right\vert ^{2}M_{\nu}\right)$.
$\boldsymbol{J}_{\nu}$ is the gauge-invariant nuclear current density.

By rewriting the conditional electronic Schr\"{o}dinger equation,
namely, Eq. (\ref{e-EOM-full}) in a more suggestive way, that's,
$i\partial_{t}\left\vert \phi\left(t,\boldsymbol{R}\right)\right\rangle =\left[H_{0}\left(\boldsymbol{R},t\right)+\lambda H_{1}\left(\boldsymbol{R},t\right)\right]\left\vert \phi\left(t,\boldsymbol{R}\right)\right\rangle $,
where $H_{0}=H^{BO}-\varepsilon^{A}$ defines the unperturbed Hamiltonian
and $H_{1}\left(\boldsymbol{R},t\right)=V^{en}\left[\phi,\chi\right]\left(\boldsymbol{R},t\right)$
represents the perturbation, the standard quantum mechanical perturbation
theory then immediately gives 
\begin{equation}
\left\vert \phi\left(t,\boldsymbol{R}\right)\right\rangle =\left\vert \phi^{\left(0\right)}\left(t,\boldsymbol{R}\right)\right\rangle +\lambda\left\vert \phi^{\left(1\right)}\left(t,\boldsymbol{R}\right)\right\rangle +\mathcal{O}\left(\lambda^{2}\right),\label{e-expand}
\end{equation}
where $\lambda$ is the usual bookkeeping parameter that will be set
to $\lambda=1$ in the end, as in standard quantum mechanical perturbation
theory. Since the order of $\lambda$ represents the perturbative
order of $V^{en}$ and since, on the other hand, $V^{en}$ is proportional
to $\mu$, the usefulness of this small parameter $\mu$ becomes immediately
apparent for the perturbation theory. Importantly, \textit{all} non-adiabatic
effects, i.e., \textit{all} contributions to $\phi$ beyond the pure
BO propagation resulting from $H_{0}$, are proportional to $\mu$.
Hence, with $\mu$ being a very small number, we expect low-order
perturbation theory in $V^{en}$ to be rather accurate. 

Perturbative methods are often used with a finite-order truncation,
i.e., $\left\vert \phi\left(t,\boldsymbol{R}\right)\right\rangle \rightarrow\left\vert \phi_{m}\left(t,\boldsymbol{R}\right)\right\rangle \equiv\sum_{l=0}^{m}\lambda^{l}\left\vert \phi^{\left(l\right)}\left(t,\boldsymbol{R}\right)\right\rangle $
with $\lambda=1$ and $m$ the highest order of interest. In our case,
the unperturbed state is simply given by the BO dynamics, 

\begin{equation}
i\hbar\partial_{t}\left\vert \phi^{\left(0\right)}\left(t,\boldsymbol{R}\right)\right\rangle =\left(H^{BO}\left(\boldsymbol{R}\right)-\varepsilon^{A}\left[\phi^{\left(0\right)}\right]\left(\boldsymbol{R},t\right)\right)\left\vert \phi^{\left(0\right)}\left(t,\boldsymbol{R}\right)\right\rangle .\label{nEXFe-p0}
\end{equation}
For many physical questions, it will be sufficient to consider only
the first-order correction given by
\begin{equation}
i\partial_{t}\left\vert \phi^{\left(1\right)}\left(t,\boldsymbol{R}\right)\right\rangle =\left[H^{BO}\left(\boldsymbol{R}\right)-\varepsilon^{A}\left[\phi^{\left(0\right)}\right]\left(\boldsymbol{R},t\right)\right]\left\vert \phi^{\left(1\right)}\left(t,\boldsymbol{R}\right)\right\rangle +V^{en}\left[\phi^{\left(0\right)},\chi\right]\left(\boldsymbol{R},t\right)\left\vert \phi^{\left(0\right)}\left(t,\boldsymbol{R}\right)\right\rangle .\label{e1-EOM}
\end{equation}
Noticeably, the corrections $\left\vert \phi^{\left(l\right)}\left(t,\boldsymbol{R}\right)\right\rangle $
with $l\ge1$ depend on the nuclear state only via the nuclear momentum
function $\boldsymbol{\mathfrak{p}}_{\nu}\left[\phi,\chi\right]\left(\boldsymbol{R},t\right)$
Eq. (\ref{pzero-q}). To carry out calculations of these corrections
in practice, one must also determine the value of $\boldsymbol{\mathfrak{p}}_{\nu}\left[\phi,\chi\right]\left(\boldsymbol{R},t\right)$
appearing there. In principle, without further approximations imposed
on the nuclear TDSE, $\boldsymbol{\mathfrak{p}}_{\nu}\left(\boldsymbol{R},t\right)$
should be determined by solving Eq. (\ref{EXF-EOM-nu}) self-consistently
with the replacement of $\left\vert \phi\left(t,\boldsymbol{R}\right)\right\rangle $
in $\varepsilon\left[\phi\right]\left(\boldsymbol{R},t\right)$ and
$\boldsymbol{A}_{\nu}\left[\phi\right]\left(\boldsymbol{R},t\right)$
by $\left\vert \phi_{m}\left(t,\boldsymbol{R}\right)\right\rangle $.
Crucially, since any additional approximations made for the nuclei
only affect the electronic factor through the value of $\boldsymbol{\mathfrak{p}}_{\nu}\left(\boldsymbol{R},t\right)$
without affecting the perturbation structure itself, one has the flexiblity
of choosing different approxmations for describing the nuclear states.
Below we will illustrate this point by appealing to two very different
approximations for the nuclear states. 

\subsubsection{Indepedent post-perturbation approximations for the nuclear subsystem}

\label{main-bodytext-ENAPT-postpert}

In the first approximation, instead of retaining the full self-consistency
in obtaining $\boldsymbol{\mathfrak{p}}_{\nu}\left(\boldsymbol{R},t\right)$
from solving Eq. (\ref{EXF-EOM-nu}) with the potentials $\varepsilon\left[\phi_{m}\right]\left(\boldsymbol{R},t\right)$
and $\boldsymbol{A}_{\nu}\left[\phi_{m}\right]\left(\boldsymbol{R},t\right)$
with $m\ge1$, one approximates the potentials simply by $\varepsilon\left[\phi^{\left(0\right)}\right]\left(\boldsymbol{R},t\right)$
and $\boldsymbol{A}_{\nu}\left[\phi^{\left(0\right)}\right]\left(\boldsymbol{R},t\right)$.
This approximation for describing the nuclear states has been used
in the practical application of EF-NVPT to the calculation of the
electronic fluxes induced by nonzero nuclear velocity \cite{Axel163316}.
Although EF-NVPT, at first glance, appears to formulate the electronic
perturbation differently, it can be reproduced from Eq. (\ref{e1-EOM})
under appropriate conditions explained below. The unperturbed electronic
state is chosen to be $\left\vert \phi^{\left(0\right)}\left(t,\boldsymbol{R}\right)\right\rangle =\left\vert \varphi_{k}\left(\boldsymbol{R}\right)\right\rangle $
and therefore $\varepsilon^{A}\left[\phi^{\left(0\right)}\right]\left(\boldsymbol{R},t\right)=\varepsilon_{k}\left(\boldsymbol{R}\right)$.
By ignoring $\varepsilon^{NA}\left[\phi^{\left(0\right)}\right]\left(\boldsymbol{R},t\right)$
and $U_{K}^{en}\left[\phi^{\left(0\right)}\right]\left(\boldsymbol{R},t\right)$
altogether in $V^{en}\left[\phi^{\left(0\right)},\chi\right]\left(\boldsymbol{R},t\right)$
and setting the left-hand side of Eq. (\ref{e1-EOM}) to zero together
with taking $\boldsymbol{A}_{\nu}\left[\varphi_{k}\right]\left(\boldsymbol{R}\right)=0$
everywhere, then Eq. (\ref{e1-EOM}) becomes the familiar Sternheimer
equation, namely, $\left[H^{BO}\left(\boldsymbol{R}\right)-\varepsilon_{k}\left(\boldsymbol{R}\right)\right]\left\vert \phi^{\left(1\right)}\left(t,\boldsymbol{R}\right)\right\rangle =\sum_{\nu}\boldsymbol{\mathfrak{p}}_{\nu}\left(\boldsymbol{R},t\right)/M_{\nu}\cdot\left[-i\boldsymbol{\nabla}_{\nu}\right]\left\vert \varphi_{k}\left(\boldsymbol{R}\right)\right\rangle $
used there. The nuclear velocity field is identified with $\boldsymbol{\mathfrak{p}}_{\nu}\left(\boldsymbol{R},t\right)/M_{\nu}$.
One can also study the problem without invoking the Sternheimer equation.
Instead, as both the unperturbed choice $\phi^{\left(0\right)}$ and
$\boldsymbol{\mathfrak{p}}_{\nu}\left(\boldsymbol{R},t\right)$ (solved
from using $\varepsilon_{\mu}\left[\phi^{\left(0\right)}\right]\left(\boldsymbol{R},t\right)$
and $\boldsymbol{A}_{\nu}\left[\phi^{\left(0\right)}\right]\left(\boldsymbol{R},t\right)$
in Eq. (\ref{EXF-EOM-nu}) as the chosen approximation for the nuclear
subsystem) are known, one is able to solve Eq. (\ref{e1-EOM}) as
an initial value problem, as mentioned above. This discussion indicates
that the same approximation made for nuclei can be suited to different
treatments of the electronic DOF, highlighting the advantages of the
EF framework in allowing separate treatments for these two fundamentally
different DOFs, electrons and nuclei. 

In the second approximation we follow the spirit of mixed-quantum-classical
approaches and treat the nuclei as classical particles. This means
to approximate the complex-valued and parametrically $\boldsymbol{R}$-dependent
nuclear momentum function $\boldsymbol{\mathfrak{p}}_{\nu}\left(\boldsymbol{R},t\right)$
by the real-valued classical momentum $\boldsymbol{P}_{\nu t}^{c}$
which has no parametric $\boldsymbol{R}$-dependence. Then $\boldsymbol{P}_{\nu t}^{c}$
along with the classical positions $\boldsymbol{R}_{\nu t}^{c}$ are
determined by the classical Newtonian EOMs, including, in principle,
the Lorentz-like forces coming from the vector potential \cite{Abedi10123002,Li2022113001}.
With such classical approximation, the Newtonian equations are self-consistently
propagated with the force-generating potentials given by $\varepsilon\left[\phi_{m=1}\right]\left(\boldsymbol{R},t\right)$
and $\boldsymbol{A}_{\nu}\left[\phi_{m=1}\right]\left(\boldsymbol{R},t\right)$.
Interestingly, a direct application of the NAPT detailed here with
the electronic factor truncated at first order \cite{Tu2025043075}
shows that with classical nuclei carrying only classical momentum,
without involving any nuclear quantum momentum, the electronic decoherence
accompanying the passing through an avoided crossing can be readily
captured by this approximation. This is unattainable by Ehrenfest
dynamics in which the nuclei are also governed by the classical Newtonian
dynamics. 

The above cases represent two very different approximations, one quantum
and one classical, for the nuclear subsystem. This illustrates the
flexibility of the framework of NAPT to incorporate independent approximations
for the nuclear subsystem, chosen for the physical purposes to be
investigated. To establish this method fully, we now turn to discuss
two most pertinent features of the EF, namely, the partial normalisation
condition and the gauge covariance \cite{Abedi10123002,Suzuki14040501,Min15073001,Li2022113001,Arribas2024233201,Abedi1222A530,Min2014263004,Requist16042108,Requist17062503}.
The exactness of the EF approach ensures that these two properties
hold as long as no further approximations are involved. Below we shall
demonstrate that these properties also hold within any finite-order
truncation of NAPT.

\subsubsection{partial normalisation condition and gauge covariance}

\label{main-bodytext-ENAPT-exactcond}

We have seen that the additional approximation used to practically
obtain $\boldsymbol{\mathfrak{p}}_{\nu}\left(\boldsymbol{R},t\right)$
can affect the details of $\left\vert \phi_{m}\left(t,\boldsymbol{R}\right)\right\rangle $.
Now we shall point out that, independently of how $\boldsymbol{\mathfrak{p}}_{\nu}\left(\boldsymbol{R},t\right)$
is obtained, $\left\vert \phi_{m}\left(t,\boldsymbol{R}\right)\right\rangle $
as the order-$m$ approximation to the exact electronic factor is
normalised to unity (at each given $\left(t,\boldsymbol{R}\right)$).
This can be seen by formally substituting the expansion, Eq. (\ref{e-expand}),
up to within order-$m$ into $\partial_{t}\left(\left\langle \phi\left(t,\boldsymbol{R}\right)\left\vert \phi\left(t,\boldsymbol{R}\right)\right.\right\rangle \right)$
with use of Eq. (\ref{e-EOM-full}) for $\partial_{t}\left\vert \phi\left(t,\boldsymbol{R}\right)\right\rangle $
and $\partial_{t}\left\langle \phi\left(t,\boldsymbol{R}\right)\right\vert $
to obtain the order-$l$ equations for $0\le l\le m$. This gives
$\partial_{t}\left[\sum_{n=0}^{l}\left\langle \phi^{\left(n\right)}\left(t,\boldsymbol{R}\right)\left\vert \phi^{\left(l-n\right)}\left(t,\boldsymbol{R}\right)\right.\right\rangle \right]=0$.
It implies $\sum_{n=0}^{l}\left\langle \phi^{\left(n\right)}\left(t,\boldsymbol{R}\right)\left\vert \phi^{\left(l-n\right)}\left(t,\boldsymbol{R}\right)\right.\right\rangle $
is a constant for all time $t$ so its value is fixed at $t=0$. Upon
the recognition that the initial state assignment to the electronic
factor is only done via the zeroth-order $\left\vert \phi^{\left(0\right)}\left(0,\boldsymbol{R}\right)\right\rangle =\left\vert \phi\left(0,\boldsymbol{R}\right)\right\rangle $
while $\left\vert \phi^{\left(l\ge1\right)}\left(0,\boldsymbol{R}\right)\right\rangle =0$,
the order-$l$ contribution to the norm is found to be $\sum_{n=0}^{l}\left\langle \phi^{\left(n\right)}\left(t,\boldsymbol{R}\right)\left\vert \phi^{\left(l-n\right)}\left(t,\boldsymbol{R}\right)\right.\right\rangle =0$
for all time $t$ and $l\ge1$.Therefore, the partial normalisation
condition is fulfilled independently of how $\boldsymbol{\mathfrak{p}}_{\nu}\left(\boldsymbol{R},t\right)$
is obtained. 

We now discuss what happens when a change of gauge is performed. The
previous discussions reveal that it is sometimes preferable to approximate
$\boldsymbol{\mathfrak{p}}_{\nu}\left(\boldsymbol{R},t\right)$ without
fully self-consistently solving Eq. (\ref{EXF-EOM-nu}) with the potentials
pinned to $\varepsilon\left[\phi_{m}\right]\left(\boldsymbol{R},t\right)$
and $\boldsymbol{A}_{\nu}\left[\phi_{m}\right]\left(\boldsymbol{R},t\right)$.
Hereby we examine the situation where we approximate the electronic
factor by $\phi_{m}$ but for the nuclei the potential energy functions
are instead approximated with $\phi_{m^{\prime}}$, allowing $m^{\prime}\le m$.
Expplicitly, this gives $\boldsymbol{A}_{\nu}\left[\phi_{m^{\prime}}\right]\left(\boldsymbol{R},t\right)=\sum_{l=0}^{m^{\prime}}\alpha^{l}\boldsymbol{A}_{\nu}^{\left(l\right)}\left[\phi\right]\left(\boldsymbol{R},t\right)$,
in which $\boldsymbol{A}_{\nu}^{\left(l\right)}\left[\phi\right]\left(\boldsymbol{R},t\right)=\sum_{n=0}^{l}\left\langle \phi^{\left(n\right)}\left(t,\boldsymbol{R}\right)\left\vert -i\boldsymbol{\nabla}_{\nu}\phi^{\left(l-n\right)}\left(t,\boldsymbol{R}\right)\right.\right\rangle $.
By an arbitrarily chosen gauge function $S\left(\boldsymbol{R},t\right)$,
we define a gauge transformation $\left\vert \bar{\phi}\left(t,\boldsymbol{R}\right)\right\rangle =e^{iS\left(\boldsymbol{R},t\right)}\left\vert \phi\left(t,\boldsymbol{R}\right)\right\rangle $
and $\bar{\chi}\left(\boldsymbol{R},t\right)=e^{-iS\left(\boldsymbol{R},t\right)}\chi\left(\boldsymbol{R},t\right)$.
Since $\left\vert \bar{\phi}_{m^{\prime}}\left(t,\boldsymbol{R}\right)\right\rangle =e^{iS\left(\boldsymbol{R},t\right)}\left\vert \phi_{m^{\prime}}\left(t,\boldsymbol{R}\right)\right\rangle $,
for each order $l\le m^{\prime}$ that leads to $\left\vert \bar{\phi}^{\left(l\right)}\left(t,\boldsymbol{R}\right)\right\rangle =e^{iS\left(\boldsymbol{R},t\right)}\left\vert \phi^{\left(l\right)}\left(t,\boldsymbol{R}\right)\right\rangle $.
Using the universal property $\sum_{n=0}^{l}\left\langle \phi^{\left(n\right)}\left(t,\boldsymbol{R}\right)\left\vert \phi^{\left(l-n\right)}\left(t,\boldsymbol{R}\right)\right.\right\rangle =0$,
obtained independently of how $\chi$ is treated, we then find 
\begin{equation}
\boldsymbol{A}_{\nu}\left[\bar{\phi}_{m^{\prime}}\right]\left(\boldsymbol{R},t\right)=\boldsymbol{A}_{\nu}\left[\phi_{m^{\prime}}\right]\left(\boldsymbol{R},t\right)+\boldsymbol{\nabla}_{\nu}S\left(\boldsymbol{R},t\right).\label{gg-v}
\end{equation}
By the same token, we also see that
\begin{equation}
\varepsilon\left[\bar{\phi}_{m^{\prime}}\right]\left(\boldsymbol{R},t\right)=\varepsilon\left[\phi_{m^{\prime}}\right]\left(\boldsymbol{R},t\right)+\partial_{t}S\left(\boldsymbol{R},t\right).\label{gg-s}
\end{equation}
Therefore the gauge transformation for the potentials evaluated using
the truncated electronic factor is formally no different from that
evaluated using the full electronic factor. The gauge invaraince of
Eq. (\ref{EXF-EOM-nu}) approximated by $\varepsilon\left[\phi_{m^{\prime}}\right]\left(\boldsymbol{R},t\right)$
and $\boldsymbol{A}_{\nu}\left[\phi_{m^{\prime}}\right]\left(\boldsymbol{R},t\right)$
as well as that of the corresponding classical forces (which consists
of the electric-like force $\left[\partial\boldsymbol{A}_{\nu}\left[\phi_{m^{\prime}}\right]\left(\boldsymbol{R},t\right)/\partial t-\partial\varepsilon\left[\phi_{m^{\prime}}\right]\left(\boldsymbol{R},t\right)/\partial\boldsymbol{R}_{\nu}\right]_{\boldsymbol{R}=\left\{ \boldsymbol{R}_{\nu t}^{c}\right\} }$
and the magnetic-like-force $\sum_{\nu^{\prime}\beta}\dot{R}_{\nu^{\prime}\beta t}^{c}\left(\partial A_{\nu\alpha}\left[\phi_{m^{\prime}}\right]\left(\boldsymbol{R},t\right)/\partial R_{\nu^{\prime}\beta}-\partial A_{\nu^{\prime}\beta}\left[\phi_{m^{\prime}}\right]\left(\boldsymbol{R},t\right)/\partial R_{\nu\alpha}\right)_{\boldsymbol{R}=\left\{ \boldsymbol{R}_{\nu t}^{c}\right\} }$)
is immediatrely seen from Eqs. (\ref{gg-v}) and (\ref{gg-s}). Here
$\dot{R}_{\nu^{\prime}\beta t}^{c}$ is the classical velocity of
nucleus $\nu^{\prime}$ in the $\beta$ direction at time $t$. Summarising
this section, we have established that the partial normalisation condition
and the gauge covariance of NAPT hold regardless which approximations
we apply to the nuclear Schr\"{o}dinger equation for obtaining the
nuclear momentum function. 

\subsection{The Berry phase of the $E\otimes e$ Jahn-Teller model}

\label{examples-BerryPhase}

Having introduced the general formulation of NAPT, the goal of the
present section is to demonstrate the practical usefulness of the
method. Previous applications of EF-NVPT often considered cases in
which the vector potential can be gauged to zero \cite{Scherrer15074106,Eich16054110,Axel163316,Scherrer17031035}.
Here, we focus on a scenario where this is not possible: the Berry
phase in Jahn-Teller systems which feature conical intersections.
In the absence of nonadiabatic electron--nuclear correlations, Berry
phases in such systems are known to be topological \cite{Longuet-Higgins19581,Herzberg6377,OBrien93688,JoubertDoriol20177365}.
However, exact treatments based on the full electron-nuclear Hamiltonian
rather the BO Hamiltonian have shown that once nonadiabatic effects
are fully accounted for (by exact numerical solutions or asymptotic
analysis of the exact nonlinear differential equations satisfied by
the EF factors), these phases become geometric \cite{Min2014263004,Requist16042108,Requist17062503,Ibele202311625,Martinazzo2024243002}.
In particular, the value of the exact geometric phase is path-dependent.
The evaluation of the exact geometric phase requires knowledge of
the exact Berry connection whose computation needs the exact electronic
factor $\left\vert \phi\left(\boldsymbol{R}\right)\right\rangle $
as input. The latter, however, is very hard to compute, especially
if one aims at an ab initio treatment. The numerical solution of the
EF equation determining the conditional electronic wavefunction is
nearly as hard as the fully correlated electronic-nuclear problem,
in some cases, maybe even harder \cite{Gossel2019154112}. From the
numerical point of view, it is therefore a highly important goal to
design approximation schemes for the solution of the EF EOM, such
as the NAPT presented above. In what follows we will demonstrate that
even the first-order NAPT correction to the zeroth order electronic
factor accurately accounts for the departure of the true geometric
phase from its adiabatic limit.

Since the traditional BO Berry phase has mainly been studied for stationary
states, we employ here the time-independent version, $\Psi\left(\boldsymbol{r},\boldsymbol{R}\right)=\chi\left(\boldsymbol{R}\right)\phi\left(\boldsymbol{r}\left\vert \boldsymbol{R}\right.\right)$,
of the EF \cite{Gidopoulos-ArX0502433,Gidopoulos201420130059}. The
latter can also be obtained from the time-dependent EOM, Eqs. (\ref{EXF-EOM-nu})
and (\ref{e-EOM-set}): we make the time-independent ansatz $\left\vert \phi\left(t,\boldsymbol{R}\right)\right\rangle =\left\vert \phi\left(\boldsymbol{R}\right)\right\rangle $
and set $\chi\left(\boldsymbol{R},t\right)=e^{-iEt}\chi\left(\boldsymbol{R}\right)$.
These lead to 
\begin{equation}
\left\{ \mu\sum_{\nu}\frac{\left[\left(-i\boldsymbol{\nabla}_{\nu}\right)+\boldsymbol{A}_{\nu}\left[\phi\right]\left(\boldsymbol{R}\right)\right]^{2}}{2M_{\nu}}+\varepsilon\left(\boldsymbol{R}\right)\right\} \chi\left(\boldsymbol{R}\right)=E\chi\left(\boldsymbol{R}\right),\label{stat-nu-eom}
\end{equation}
where $E$ is the total energy of the full electron-nuclear correlated
system and 
\begin{equation}
\left[H^{BO}\left(\boldsymbol{R}\right)+\mu U^{en}\left[\phi,\chi\right]\left(\boldsymbol{R}\right)\right]\left\vert \phi\left(\boldsymbol{R}\right)\right\rangle =\varepsilon\left(\boldsymbol{R}\right)\left\vert \phi\left(\boldsymbol{R}\right)\right\rangle .\label{stat-e-eom}
\end{equation}
The first-order correction to the electronic factor according to the
above equation then becomes
\begin{equation}
\left\vert \phi^{\left(1\right)}\left(\boldsymbol{R}\right)\right\rangle =-\left[H^{BO}\left(\boldsymbol{R}\right)-\varepsilon^{\left(0\right)}\left(\boldsymbol{R}\right)\right]^{-1}V^{en}\left[\phi^{\left(0\right)},\chi\right]\left(\boldsymbol{R}\right)\left\vert \phi^{\left(0\right)}\left(\boldsymbol{R}\right)\right\rangle ,\label{Sternheimer-s1}
\end{equation}
where $\varepsilon^{\left(0\right)}\left(\boldsymbol{R}\right)=\left\langle \phi^{\left(0\right)}\left(\boldsymbol{R}\right)\left\vert H^{BO}\left(\boldsymbol{R}\right)\right\vert \phi^{\left(0\right)}\left(\boldsymbol{R}\right)\right\rangle $.
Using Eq. (\ref{Sternheimer-s1}) the corrected Berry connection is
readily calculated as $\boldsymbol{A}\left[\phi\right]\left(\boldsymbol{R}\right)=\boldsymbol{A}^{\left(0\right)}\left(\boldsymbol{R}\right)+\lambda\boldsymbol{A}^{\left(1\right)}\left(\boldsymbol{R}\right)+\mathcal{O}\left(\lambda^{2}\right)$,
where $\boldsymbol{A}^{\left(0\right)}\left(\boldsymbol{R}\right)=\left\langle \phi^{\left(0\right)}\left(\boldsymbol{R}\right)\left\vert -i\boldsymbol{\nabla}\phi^{\left(0\right)}\left(\boldsymbol{R}\right)\right.\right\rangle $,
$\boldsymbol{A}^{\left(1\right)}\left(\boldsymbol{R}\right)=\left\langle \phi^{\left(1\right)}\left(\boldsymbol{R}\right)\left\vert -i\boldsymbol{\nabla}\phi^{\left(0\right)}\left(\boldsymbol{R}\right)\right.\right\rangle +\text{c.c.}$
Likewise the corrected Berry phase takes the form, $\gamma\left(Q\right)=\gamma^{\left(0\right)}\left(Q\right)+\lambda\gamma^{\left(1\right)}\left(Q\right)+\mathcal{O}\left(\lambda^{2}\right)$
with $\gamma^{\left(n\right)}\left(Q\right)=\oint_{\mathcal{C}\left(Q\right)}\text{d}\boldsymbol{R}\cdot\boldsymbol{A}^{\left(n\right)}\left(\boldsymbol{R}\right)$
for $n=0,1$. The BO Berry phase $\gamma^{BO}\left(Q\right)$ is defined
to be $\gamma^{\left(0\right)}\left(Q\right)$ with $\left\vert \phi^{\left(0\right)}\left(\boldsymbol{R}\right)\right\rangle $
given by the lowest energy eigenstate of $H^{BO}\left(\boldsymbol{R}\right)$.
The main focus of the present application is $\gamma^{\left(1\right)}\left(Q\right)$,
which represents the minimal extent of nonadiabatic treatment within
NAPT. Below we first recall the Berry phase properties in the adiabatic
limit and then we discuss the nonadiabatic corrections to it. 

\subsubsection{Berry phase of the $E\otimes e$ Jahn-Teller model in the adiabatic
limit}

\label{examples-BerryPhase-adiabatic}

The $E\otimes e$ Jahn-Teller system provides a minimal model exhibiting
a conical intersection, which in the adiabatic limit leads to a topological
Berry phase \cite{Longuet-Higgins19581,Herzberg6377,OBrien93688,JoubertDoriol20177365}.
Its BO Hamiltonian is
\begin{equation}
H^{JT,BO}\left(\boldsymbol{R}\right)=\left(K/2\right)Q^{2}+gQ\left(\boldsymbol{\sigma}\times\hat{\boldsymbol{r}}\right)_{2}.\label{JTBO-H}
\end{equation}
Here $\boldsymbol{R}=\left(R_{1},R_{3}\right)$ describes the displacement
coordinates of two nuclear vibrational normal modes with $\boldsymbol{R}=0$
corresponding to the equilibrium configuration. We have used the polar
coordinate system $\boldsymbol{R}=Q\left(\cos\theta,\sin\theta\right)$
with $Q\equiv\sqrt{R_{1}^{2}+R_{3}^{2}}$ the radial length, $\theta$
the polar angle and $\hat{\boldsymbol{r}}=\left(\cos\theta,0,\sin\theta\right)$
the directional unit vector. $\boldsymbol{\sigma}=\left(\sigma_{1},\sigma_{2},\sigma_{3}\right)$
is made of the Pauli matrices built from two diabatic states $\left\vert \pm\right\rangle $,
satisfying $\sigma_{3}\left\vert \pm\right\rangle =\pm\left\vert \pm\right\rangle $,
that define the electronic Hilbert space for this problem. $g>0$
and $K>0$ are parameters of the model. The BO Hamiltonian Eq. (\ref{JTBO-H})
at each $\boldsymbol{R}=\left(Q,\theta\right)$ can be diagonalised,
namely, $H^{JT,BO}\left(\boldsymbol{R}\right)\left\vert \varphi_{\pm}\left(\boldsymbol{R}\right)\right\rangle =\varepsilon_{\pm}\left(\boldsymbol{R}\right)\left\vert \varphi_{\pm}\left(\boldsymbol{R}\right)\right\rangle $,
with the eigenenergies $\varepsilon_{\pm}\left(\boldsymbol{R}\right)=\left(K/2\right)Q^{2}\pm gQ$.
One immediately realises that the electronic energies are degenerate
at zero displacement $Q=0$ and a gap opens in proportion to $Q$.
We calculate the BO Berry phase, $\gamma^{BO}\left(Q\right)=\oint_{\mathcal{C}\left(Q\right)}\text{d}\boldsymbol{R}\cdot\boldsymbol{A}\left[\varphi_{-}\right]\left(\boldsymbol{R}\right),$
along a circular contour $\mathcal{C}\left(Q\right)$ of radius $Q$.
It has the value $\gamma^{BO}=\pi$, independently of the radius $Q$.
That the Berry phase is independent of the loop radius shows its topological
nature \cite{Longuet-Higgins19581,Herzberg6377,OBrien93688,JoubertDoriol20177365}. 

This topological character is directly related to the conical intersection
at $Q=0$ where the BO electronic states are degenerate so that the
BO Berry curvature diverges. Consequently, the BO Berry phase does
not vanish in the $Q\rightarrow0$ limit, despite the area enclosed
by $\mathcal{C}\left(Q\right)$ shrinking to zero. By contrast, the
Berry phase assoicated with the exact electronic factor is very different
\cite{Min2014263004,Requist16042108,Requist17062503}. The exact electronic
factor cannot generally be gauged to remove its $Q$-dependence \cite{Requist17062503}
and, unlike the BO eigenstates, is well-defined even at the places
where the BO surfaces have a conical intersection. In what follows,
we tackle this problem using the first-order NAPT, minimally engaging
the electron-nuclear nonadiabatic correlation by taking $\left\vert \phi^{\left(0\right)}\left(\boldsymbol{R}\right)\right\rangle =\left\vert \varphi_{-}\left(\boldsymbol{R}\right)\right\rangle $
and approximating the nuclear momentum function $\boldsymbol{\mathfrak{p}}_{\nu}\left(\boldsymbol{R}\right)$
by setting the potential energies in the eigenvalue equation for $\chi\left(\boldsymbol{R}\right)$
using $\phi^{\left(0\right)}$ alone. We treat separately the cases
of finite $Q$ and the limit $Q\rightarrow0$.

\subsubsection{Berry phase of the $E\otimes e$ Jahn-Teller model in first-order
NAPT}

\label{examples-BerryPhase-devadiabatic}

We first consider the case of finite $Q$ in $\mathcal{C}\left(Q\right)$
(details of the derivation are in Appendix \ref{BerryPhaseDetails-finiteQ}).
The approximate $\boldsymbol{\mathfrak{p}}_{\nu}\left(\boldsymbol{R}\right)$
reads 
\begin{equation}
\boldsymbol{\mathfrak{p}}\left(Q,\theta\right)=\left(\hat{\boldsymbol{\theta}}-i\hat{\boldsymbol{r}}\right)\frac{m}{Q},\label{BerryNucMomFunc-1}
\end{equation}
where $\hat{\boldsymbol{\theta}}$ is the 2D unit vector in the angular
direction orthogonal to $\hat{\boldsymbol{r}}$. Here $m$ is the
angular momentum quantum number for the nuclei. Noticeably, this result
Eq. (\ref{BerryNucMomFunc-1}) is obtained without the need to fully
access the explicit form of the nuclear density. The subsequent correction
to the electronic factor then reads $\left\vert \phi^{\left(1\right)}\left(\boldsymbol{R}\right)\right\rangle =i\left[\mu m/\left(4gM_{0}Q^{3}\right)\right]\left[\cos\left(\theta/2\right)\left\vert +\right\rangle -\sin\left(\theta/2\right)\left\vert -\right\rangle \right]$.
Here $M_{0}$ is the mass carried by the two normal modes. Recall
that in the adiabatic limit, while the first term in $H^{JT,BO}\left(\boldsymbol{R}\right)$
plays no role in its eigenstates that determine the Berry phase, the
second term $gQ\left(\boldsymbol{\sigma}\times\hat{\boldsymbol{r}}\right)_{2}$
dictates that the BO eigenstate $\left\vert \varphi_{-}\left(\boldsymbol{R}\right)\right\rangle $
is readily independent of $Q$, implying the $Q$-independence of
the BO Berry phase. Interestingly, by mimimally extending $\left\vert \varphi_{-}\left(\boldsymbol{R}\right)\right\rangle $
to $\left\vert \varphi_{-}\left(\boldsymbol{R}\right)\right\rangle +\left\vert \phi^{\left(1\right)}\left(\boldsymbol{R}\right)\right\rangle $,
we see that the $Q$-dependence in $\left\vert \phi^{\left(1\right)}\left(\boldsymbol{R}\right)\right\rangle $
cannot be gauged away. The result for the geometric phase reads
\begin{equation}
\gamma^{\left(1\right)}\left(Q\right)=-\frac{\mu m\pi}{2gM_{0}Q^{3}}.\label{BerryPhase-ct1}
\end{equation}
It is then evident that the path-independent topological Berry phase
$\gamma^{\left(0\right)}\left(Q\right)=\pi$ is modified by Eq. (\ref{BerryPhase-ct1})
due to electron--nuclear correlation, acquiring a geometric character
whereby its value depends on the loop radius $Q$. We note in passing
the same result can be extracted from the asymptotic analysis of the
fully exact electronic factor $\left\vert \phi\left(Q,\theta\right)\right\rangle $
in Ref. \cite{Requist17062503} (see Appendix \ref{BerryPhaseDetails-finiteQ}).

\begin{figure}[h] \includegraphics[width=17cm, height=4.5 cm]{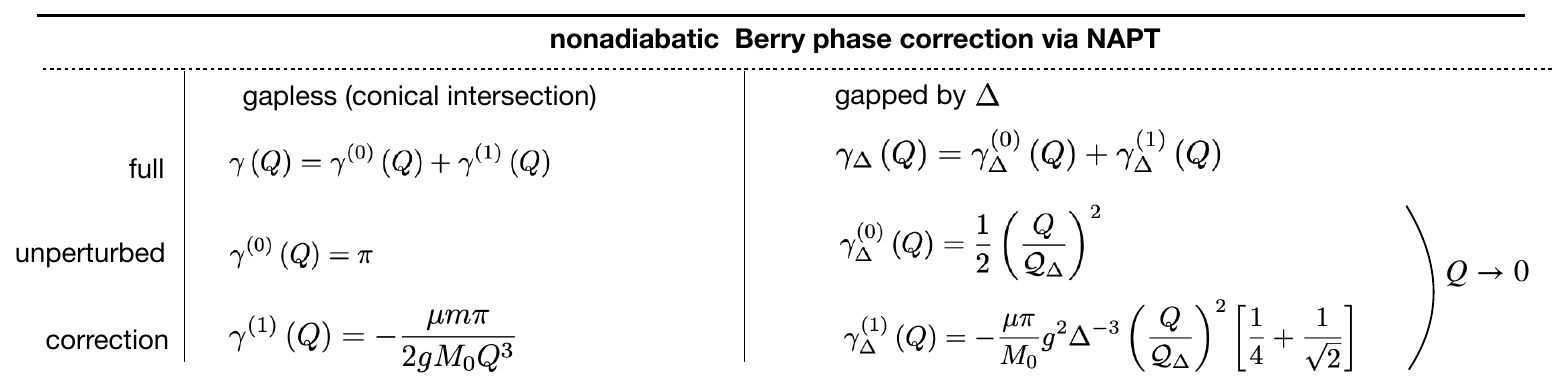} 
\caption{The main findings of applying NAPT developed here to the Berry phase problem in Jahn-Teller systems for degenerate (left panel) and gapped (right panel) cases. } 
\label{BerryCorr} 
\end{figure}

We now turn to the limit $Q\rightarrow0$. Naively taking $Q=0$ in
$H^{JT,BO}\left(\boldsymbol{R}\right)$ for the BO eigenvalue equation
does not give us a definite BO eigenstate to serve as the unperturbed
state. Since a perturbative method presupposes a well-defined unperturbed
state, to proceed, we regulate the BO Hamiltonian by adding to it
a gap-opening term, namely, $H^{JT,BO}\left(\boldsymbol{R}\right)\rightarrow H_{\Delta}^{JT,BO}\left(\boldsymbol{R}\right)=H^{JT,BO}\left(\boldsymbol{R}\right)+\Delta\sigma_{2}$
with $\Delta>0$ being a constant such that its eigenstates are well-defined
as $Q\rightarrow0$. For clarity, we use the same notation for quantities
calculated before with $\Delta=0$ but we will add a superscript or
subscript $\Delta$ to the symbol in order to make the distinction
explicit. We leave the details of the derivation for this part in
Appendix \ref{BerryPhaseDetails-smallQ}. To analyse the vicinity
of $Q\to0$, we define a length $\mathcal{Q}_{\Delta}=\Delta/g$ .
One then finds to the lowest order in $Q/\mathcal{Q}_{\Delta}$ that
$\gamma_{\Delta}^{\left(0\right)}\left(Q\right)=\left(\pi/2\right)\left(Q/\mathcal{Q}_{\Delta}\right)^{2}$.
The nuclear momentum function $\boldsymbol{\mathfrak{p}}\left(Q,\theta\right)$
remains approximately given by Eq. (\ref{BerryNucMomFunc-1}) in the
vicinity of $Q/\mathcal{Q}_{\Delta}\ll1$. The leading contribution
in the order of $Q/\mathcal{Q}_{\Delta}$ to the Berry phase correction
turns out to be 
\begin{equation}
\gamma_{\Delta}^{\left(1\right)}\left(Q\right)=-\frac{\mu\pi}{M_{0}}g^{2}\Delta^{-3}\left(\frac{Q}{\mathcal{Q}_{\Delta}}\right)^{2}\left[\frac{1}{4}+\frac{1}{\sqrt{2}}\right].\label{BerryPhase-Dct1}
\end{equation}
Note that the order $\mathcal{O}\left(\left(Q/\mathcal{Q}_{\Delta}\right)^{0}\right)$
contribution to $\gamma_{\Delta}^{\left(1\right)}\left(Q\right)$,
carried with $\boldsymbol{\mathfrak{p}}\left(Q,\theta\right)$, is
cancelled so the result becomes independent of the angular momentum
quantum number $m$. Noticeably, Eq. (\ref{BerryPhase-Dct1}) approaches
a vanishing Berry phase, $\lim_{Q\rightarrow0}\gamma_{\Delta}^{\left(1\right)}\left(Q\right)=0$,
with $\gamma_{\Delta}^{\left(1\right)}\left(Q\right)\propto Q^{2}$.
Putting the BO Berry phase together with its nonadiabatic correction,
here we obtan the quadratic scaling $\gamma_{\Delta}\left(Q\right)\propto Q^{2}$
for $Q\rightarrow0$. This is consistent with the result of Ref. \cite{Requist17062503},
where the fully exact electronic factor was analysed using the singular
BO Hamiltonian. There, the exact treatment also yields $\gamma^{\text{exact}}\left(Q\right)\propto Q^{2}$
as $Q\rightarrow0$ (see again Appendix \ref{BerryPhaseDetails-smallQ}).
These results are summarised in Fig. \ref{BerryCorr}.

\section{discussion and outlook}

\label{conclu}

The power of the EF approach lies in the separation of the fully correlated
electron-nuclear Schr\"{o}dinger equation into two formally exact
equations of motion, one determining a purely nuclear wave function
$\chi\left(\boldsymbol{R},t\right)$, the other yielding a many-electron
amplitude $\phi\left(\boldsymbol{r}\left\vert t,\boldsymbol{R}\right.\right)$
which conditionally depends on the nuclear coordinates and on time.
This separation into two equations allows one to treat the electronic
and nuclear degrees of freedom by different methodologies, tailored
to the very different physical nature of the two species. The nuclear
equation, being a normal time-dependent Schr\"{o}dinger equation,
can be treated with standard approaches. For example, at low temperature,
when the nuclei only perform small displacements from their equilibrium
positions, the natural description of the nuclear wave function is
in terms of phonon modes. If, on the other hand, the nuclei perform
large-amplitude motion, e.g. in a scattering or dissociation process,
then single- or multiple-trajectory methods provide a good way to
solve the nuclear equation. 

The equation of motion of the electronic factor $\phi\left(\boldsymbol{r}\left\vert t,\boldsymbol{R}\right.\right)$,
on the other hand, is not a standard Schr\"{o}dinger equation due
to the appearance of first- and second-order derivatives with respect
to the nuclear coordinates. The presence of these terms mediates a
non-unitary propagation in the electronic $N$-body Hilbert space.
The numerically exact solution of the EF electronic equation of motion
is hard, especially if one aims at an ab-initio treatment. In some
cases, the exact numerical solution may even be harder than the numerical
solution of the full electron-nuclear Schr\"{o}dinger equation \cite{Gossel2019154112}.
From the numerical point of view it is therefore a highly important
goal to design efficient approximation schemes for the electronic
EOM. In this article we present a systematic perturbative approach,
NAPT, to tackle the EF electronic equation of motion. The formulation
of NAPT is motivated by the presence of a small dimensionless parameter,
the electronic-to-nuclear mass ratio $\mu$, which multiplies all
terms beyond the BO Hamiltonian. By choosing as unperturbed system
the standard BO clamped-nuclei electronic-structure problem, and by
treating the terms proportional to $\mu$ (i.e. all non-adiabatic
terms in the electronic equation) as perturbation, a systematic perturbative
expansion is obtained where the orders of perturbation theory go hand
in hand with integer powers of $\mu$. We emphasise that this perturbative
approach is applied only to the electronic equation of motion of the
EF. Usually, it is not desirable to treat the nuclear degrees of freedom
in the same way because, in the nuclear equation of motion, the parameter
$\mu$ multiplies the nuclear kinetic energy. If one were to treat
this term as perturbation, then, in the unperturbed ($\mu\rightarrow0$)
limit, the nuclei would \textquotedblleft get stuck\textquotedblright{}
at their equilibrium positions, implying that the nuclear wave function
would reduce to a product of delta functions centered at the nuclear
equilibrium positions. Clearly this would not be a very suitable zero-order
wave function to improve upon by finite orders of perturbation theory.

The perturbative approach of NAPT is specifically tailored to deal
with the electronic equation of motion. The resulting electronic wave
function is then combined with a nuclear wave function obtained in
a different way (e.g. by using trajectories or phonons). However,
since the nuclear wave function enters the electronic equation of
motion through the nuclear momentum function, Eq. (\ref{pzero-q}),
it is crucial to prove that NAPT is generally applicable, independent
of the way we treat the nuclear degrees of freedom. The exact equations
and their solutions have two essential features: gauge covariance
and the partial normalization condition of the electronic factor.
In this article we have demonstrated that these two crucial features
are preserved by any finite-order truncation of the NAPT perturbation
expansion, regardless of how the nuclear factor present in the electronic
EOM is approximated (as demonstrated in Sec. \ref{main-bodytext-ENAPT}). 

However, the presence of the nuclear factor in the electronic EOM
implies that there is an additional $\mu$-dependence entering the
equation through the nuclear momentum function Eq. (\ref{pzero-q}).
No general statements can be made on the $\mu$-dependence of this
term. The form of the $\mu$-dependence is governed by the nature
of the physical processes at hand. For the very specific case of small-amplitude
nuclear motion, the form of the $\mu$-dependence can be determined
as follows: Restricting the nuclear motion to small displacements,
$\boldsymbol{U}=\boldsymbol{R}-\boldsymbol{R}_{0}$, of the nuclei
from their equilibrium positions $\boldsymbol{R}_{0}$, we can expand
the potentials Eqs. (\ref{EXF-pot-scalsum}) and (\ref{EXF-pot-vec})
in a Tailor series in powers of $\boldsymbol{U}$ around $\boldsymbol{U}=0$.
By truncating the series of the vector potential at first order and
the series of the scalar potential at second order, and by diagonalising
the resulting Hessian, the nuclear EOM (in the static case) reduces
to a set of standard harmonic oscillator equations for the nuclear
normal modes $Q_{\alpha}\left(\boldsymbol{R}\right)$. The nuclear
density then becomes a product of normalised Gaussians: 
\begin{equation}
\left\vert \chi\left(\boldsymbol{R}\right)\right\vert ^{2}=G_{1}\left(Q_{1}\right)G_{2}\left(Q_{2}\right)G_{3}\left(Q_{3}\right)\cdots\label{HA-nucDs}
\end{equation}
with
\begin{equation}
G_{\alpha}\left(Q\right)=\frac{1}{\sigma_{\alpha}\sqrt{2\pi}}e^{-\left(Q/\sigma_{\alpha}\right)^{2}/2}.\label{HA-Gauss}
\end{equation}
The width, $\sigma_{\alpha}$, of the Gaussian $G_{\alpha}\left(Q\right)$
is
\begin{equation}
\sigma_{\alpha}=\left(\frac{\mu}{K_{\alpha}}\right)^{1/4},\label{HA-widthGS}
\end{equation}
with a spring constant $K_{\alpha}$. From Eqs. (\ref{HA-Gauss})
and (\ref{HA-widthGS}), one immediately verifies that 
\begin{equation}
\frac{\boldsymbol{\nabla}\chi}{\chi}\sim\mu^{-1/2}\label{HA-pmt}
\end{equation}
for small $\mu$.Taking into account the overall prefactor $\mu$,
this term becomes proportional to $\mu^{1/2}$ and, hence, becomes
the dominant term among all contributions to what we have defined
as perturbation in NAPT. We note that Eq. (\ref{HA-pmt}) yields also
the dominant contribution to the terms chosen as perturbation in EF-NVPT.
Hence, we expect the results NAPT and of EF-NVPT to be very similar
whenever the nuclei are constrained to small-amplitude motion. Moreover,
Eq. (\ref{HA-widthGS}) implies that the characteristic length scale
of nuclear motion, when the latter is constrained to small amplitudes,
is proportional to $\mu^{1/4}$, i.e. the small displacements of the
nuclei from their equilibrium positions must satisfy
\begin{equation}
\boldsymbol{U}\sim\mu^{1/4}.\label{HA-dis}
\end{equation}
Expanding all $\boldsymbol{R}$-dependent quantities on the r.h.s.
of the EF electronic EOM in powers of $\boldsymbol{U}$, we can identify
three sources of $\mu$-dependence in the terms defined as perturbation
in NAPT: (i) the prefactor $\mu$ multiplying all perturbative terms,
(ii) the $\boldsymbol{\nabla}\chi/\chi$ term which is proportional
to $\mu^{-1/2}$, and (iii) the powers in $\boldsymbol{U}$ which
correspond to powers of $\mu^{1/4}$. Taking all these $\mu$-dependencies
into account, a consistent theory in terms of (fractional) powers
of $\mu$ was recently formulated for the electron-phonon interactions
in solids \cite{Cohen2025075102}.

The simplifications given by Eqs. (\ref{HA-nucDs}-\ref{HA-dis})
are possible only for small-amplitude nuclear motion. We emphasise
that the perturbative approach of NAPT, as presented in this article,
is not limited to the case of small displacements of the nuclei. This
is demonstrated for the particularly delicate case of the exact molecular
geometric phase where the value of the phase is path dependent and,
hence, requires also evaluations along large nuclear paths. We demonstrate
that even the first-order-NAPT correction to the zero-order BO electronic
factor accurately accounts for the departure of the exact geometric
phase from its adiabatic BO limit. The latter, in the presence of
a conical intersection, is a quantised phase (being equal to integer
multiples of $\pi$), while the exact molecular phase is a geometric
phase showing an inverse cubic dependence on the radius of the contour.
By regularising the conical intersection with a constant gap, the
first-order-NAPT treatment yields the expected vanishing of the Berry
phase with quadratic scaling as the radius of the contour shrinks.
Both results reproduce the characteristic behaviour known from the
exact treatment of Ref. \cite{Requist17062503}. Importantly, NAPT
achieves this without invoking the full nuclear wave function, but
simply from an approximate nuclear momentum function coming from a
single BO surface..

Electron-nuclear correlation beyond the BO limit is central to many
physical and chemical phenomena. Yet, solving the fully coupled Schr\"{o}dinger
equation of Coulomb-interacting electrons and nuclei is practically
impossible, except for simple model systems. The EF-based perturbative
approach of NAPT proposed in this article represents a transparent
and structurally consistent way of tackling electron-nuclear correlations.
The successful application to the delicate problem of evaluating non-adiabatic
corrections to the traditional BO molecular Berry phase suggests the
approach may be usefully applied to many other non-adiabatic phenomena.

\appendix

\section{Details of Berry-phase correction calculations}

\label{BerryPhaseDetails}

\subsection{Case of finite $Q$}

\label{BerryPhaseDetails-finiteQ}

The BO electronic state reads $\left\vert \varphi_{-}\left(\theta\right)\right\rangle =\sin\left(\theta/2\right)\left\vert +\right\rangle +\cos\left(\theta/2\right)\left\vert -\right\rangle $
and we find the nuclear Schr\"{o}dinger equation is separable in
$\theta$ and $Q$ from which we obtain $\chi\left(Q,\theta\right)=e^{im\theta}G\left(Q\right)$,
with $m$ being the angular momentum quantum number. Without a need
to know explicitly the radial wavefunction $G\left(Q\right)$, we
can readily obtain Eq. (\ref{BerryNucMomFunc-1}). One can obtain
the same nuclear momentum function in any other gauges. The corresponding
correction to the vector potential is $\boldsymbol{A}^{\left(1\right)}\left(\boldsymbol{R}\right)=-\mu m/\left(4gM_{0}Q^{4}\right)$.
This also gives $\varepsilon_{\mu}\left[\varphi_{-}\right]\left(\boldsymbol{R}\right)=\left(K/2\right)Q^{2}-gQ+\mu\varepsilon_{-}^{na}\left(Q\right)$
where $\varepsilon_{-}^{na}\left(Q\right)=\varepsilon^{na}\left[\varphi_{-}\right]\left(\boldsymbol{R}\right)=1/\left(8M_{0}Q^{2}\right)$
which is found to be gauge-independent. With $\boldsymbol{\mathfrak{p}}\left(Q,\theta\right)$
ready, one can proceed to use Eq. (\ref{Sternheimer-s1}) to calculate
the perturbation correction to the electronic factor. Subsequently
with some algebra, we obtain These results straightforwardly yield
Eq. (\ref{BerryPhase-ct1}).

We now re-derive the above result by revisiting the exact treament
of Ref. \cite{Requist17062503}. There, the full exact electronic
factor is parameterised via $\left\vert \phi\left(Q,\theta\right)\right\rangle =\cos\left(\frac{\vartheta\left(Q\right)}{2}\right)\left\vert +\right\rangle +e^{i\theta}\sin\left(\frac{\vartheta\left(Q\right)}{2}\right)\left\vert -\right\rangle $
in which $\vartheta\left(Q\right)$ satisfies a nonlinear differential
equation that couples to the nuclear wavefunction $\chi$. By defining
a dimensionless parameter $\epsilon\equiv\mu^{1/2}K^{3/2}/\left(g^{2}M_{0}^{1/2}\right)$,
which is indeed a small parameter by the smallness of $\mu$, they
found for small enough $\epsilon$, one can approximate $\vartheta\left(Q\right)=\tan^{-1}\left(4\left(Q/Q_{0}\right)^{3}\epsilon^{-2}\right)$
where $Q_{0}=g/K$ for $Q/Q_{0}>0.3$. The parameterisation of the
conditonal electronic state at the same gives the exact Berry phase
in form of $\gamma^{\text{exact}}\left(Q\right)=\pi\left(1-\cos\left(\vartheta\left(Q\right)\right)\right)$
from which one extracts the deviation from the BO limit by $\delta\gamma\left(Q\right)=\pi-\gamma^{\text{exact}}\left(Q\right)$.
By expanding $\cos\left(\vartheta\left(Q\right)\right)$ in powers
of $\epsilon^{2}$, we find $\delta\gamma\left(Q\right)=-\pi\epsilon^{2}/\left[4\left(Q/Q_{0}\right)^{3}\right]+\mathcal{O}\left(\left(\epsilon^{2}\right)^{2}\right)$.
Restoring the definition of $\epsilon$ and neglecting $\mathcal{O}\left(\left(\epsilon^{2}\right)^{2}\right)$
for small enough $\epsilon$, we then find $\delta\gamma\left(Q\right)=\gamma^{\left(1\right)}\left(Q\right)$
of Eq. (\ref{BerryPhase-ct1}) upon taking the quantum number for
the circulating current to be $m=1/2$, which agrees with the choice
of the angular momentum quantum number to be $1/2$ in Ref. \cite{Requist17062503}. 

\subsection{Limit of $Q\rightarrow0$}

\label{BerryPhaseDetails-smallQ}

With a gap $\Delta$ opening at $Q=0$, the eigenstate of $H_{\Delta}^{JT,BO}\left(\boldsymbol{R}\right)$
that will be taken as our unperturbed starting point is parameterised
by both $Q$ and $\theta$ as $\left\vert \varphi_{-}^{\Delta}\left(\boldsymbol{R}\right)\right\rangle =e^{i\theta}\sin\left[\frac{\Theta\left(Q\right)}{2}\right]\left\vert +y\right\rangle -\cos\left[\frac{\Theta\left(Q\right)}{2}\right]\left\vert -y\right\rangle $
where $\cos\Theta\left(Q\right)=\Delta/\sqrt{\left(gQ\right)^{2}+\Delta^{2}}$
and $\sigma_{2}\left\vert \pm y\right\rangle =\pm\left\vert \pm y\right\rangle $
with the BO eigenenergy $\varepsilon_{-}^{\Delta}\left(\boldsymbol{R}\right)=\left(K/2\right)Q^{2}-g\sqrt{\left(gQ\right)^{2}+\Delta^{2}}$.
It's straightforward to see that $\gamma_{\Delta}^{\left(0\right)}\left(Q\right)=\pi\left[1-\cos\Theta\left(Q\right)\right]$
and thus its quadratic scaling for $Q/\mathcal{Q}_{\Delta}\ll1$.
The diagonal correction energy then reads $\varepsilon^{na}\left[\varphi_{-}^{\Delta}\right]\left(\boldsymbol{R}\right)=\frac{1}{2M_{0}}\left\{ -\frac{\left(1-\cos\Theta\right)^{2}}{4Q^{2}}+\frac{1}{2}\left[\left(\frac{\partial\cos\Theta}{\partial Q}\right)^{2}\frac{1}{1-\cos^{2}\Theta}+\frac{1-\cos\Theta}{Q^{2}}\right]\right\} $.
By taking $\Delta=0\rightarrow\cos\Theta\left(Q\right)=0$, one immediately
see it reduces to the previous result $\varepsilon^{na}\left[\varphi_{-}\right]\left(\boldsymbol{R}\right)=1/\left(8M_{0}Q^{2}\right)$
calculated readily with $\Delta=0$. Noticeably $\varepsilon^{na}\left[\varphi_{-}^{\Delta}\right]$
with $\Delta\ne0$ also only depends on $Q$ but not $\theta$ so
the entire scalar potential energy $\varepsilon_{\mu}\left[\varphi_{-}^{\Delta}\right]\left(\boldsymbol{R}\right)=\varepsilon_{-}^{\Delta}\left(\boldsymbol{R}\right)+\mu\varepsilon_{na}\left[\varphi_{-}^{\Delta}\right]\left(\boldsymbol{R}\right)=\varepsilon_{\mu}\left[\varphi_{-}^{\Delta}\right]\left(Q\right)$
is a function of $Q$ only. Expanding $\varepsilon_{\mu}\left[\varphi_{-}^{\Delta}\right]\left(Q\right)$
in $Q/\mathcal{Q}_{\Delta}$ yields $\varepsilon_{\mu}\left[\varphi_{-}^{\Delta}\right]\left(Q\right)\sim\mathcal{O}\left(1\right)+\mathcal{O}\left(\left(Q/\mathcal{Q}_{\Delta}\right)^{2}\right)$.
The vector potential $\boldsymbol{A}\left[\varphi_{-}^{\Delta}\right]\left(\boldsymbol{R}\right)=\hat{\boldsymbol{\theta}}A_{\theta}^{\left(0\right)}\left(Q\right)$
is also found with zero radial component and the non-vanishing angular
component is given by $A_{\theta}^{\left(0\right)}\left(Q\right)=\left(1/\left(2Q\right)\right)\left(1-\cos\Theta\right)$
as a function of $Q$ only. Substituting these results into the nuclear
Schr\"{o}dinger equation and zooming into the regime $Q/\mathcal{Q}_{\Delta}\ll1$
such that the regular terms of the scalar potential become unimportant
in comparison to terms of the form $Q^{-1}$ or $Q^{-2}$ in the kinetic
energy, the nuclear wavefunction becomes dominated by $-\frac{\mu}{2M_{0}}\left[\frac{1}{Q}\frac{\partial}{\partial Q}+\left(\frac{\partial}{\partial Q}\right)^{2}+\frac{1}{Q^{2}}\left(\frac{\partial}{\partial\theta}\right)^{2}\right]\chi\left(\boldsymbol{R}\right)\approx E_{M}\chi\left(\boldsymbol{R}\right)$.
Upon realising that $\chi\sim e^{i\theta m}Q^{m}$ and the contribution
to $\boldsymbol{\mathfrak{p}}\left(Q,\theta\right)$ from $\boldsymbol{A}\left[\varphi_{-}^{\Delta}\right]\left(\boldsymbol{R}\right)$
becomes negligible in comparison to that from $-i\boldsymbol{\nabla}\chi/\chi$
for small $Q$, one then still obtains Eq. (\ref{BerryNucMomFunc-1}).
Putting these results together then yield Eq. (\ref{BerryPhase-Dct1}).

Recall that $\gamma^{\text{exact}}\left(Q\right)=\pi\left(1-\cos\left(\vartheta\left(Q\right)\right)\right)$
for arbitrary $Q$ while for the vicinity $Q\rightarrow0$ (distinct
from $Q/Q_{0}>0.3$ used before) Ref. \cite{Requist17062503} also
showed that $\vartheta\left(Q\right)\propto Q$ . Therefore one deduces
$\gamma^{\text{exact}}\left(Q\right)\propto Q^{2}$ for $Q\rightarrow0$.

\section*{Acknowledgement}

This project has received funding from the European Research Council
(ERC) under the European Union's Horizon 2020 research and innovation
programme (grant agreement No. ERC-2017-AdG-788890). E.K.U.G. acknowledges
support as Mercator fellow within SFB 1242 at the University Duisburg-Essen.

\bibliographystyle{myunsrt} 
\bibliography{refs_bibexd-1} 

\begin{thebibliography}{10}

\bibitem{Born27457}
M.~Born and R.~Oppenheimer.
\newblock Zur quantentheorie der molekeln.
\newblock {\em Annalen der Physik}, 389(20):457--484, 1927.

\bibitem{Hagedorn86571}
George~A. Hagedorn.
\newblock High order corrections to the time-dependent born-oppenheimer
  approximation i: Smooth potentials.
\newblock {\em Annals of Mathematics}, 124(3):571--590, 1986.

\bibitem{Nafie925687}
Laurence~A. Nafie.
\newblock Velocity-gauge formalism in the theory of vibrational circular
  dichroism and infrared absorption.
\newblock {\em The Journal of Chemical Physics}, 96(8):5687--5702, 04 1992.

\bibitem{Scherrer135305}
A.~Scherrer, R.~Vuilleumier, and D.~Sebastiani.
\newblock Nuclear velocity perturbation theory of vibrational circular
  dichroism.
\newblock {\em Journal of Chemical Theory and Computation}, 9(12):5305--5312,
  2013, arXiv:https://doi.org/10.1021/ct400700c.
\newblock PMID: 26592268.

\bibitem{Scherrer15074106}
Arne Scherrer, Federica Agostini, Daniel Sebastiani, E.~K.~U. Gross, and
  Rodolphe Vuilleumier.
\newblock {Nuclear velocity perturbation theory for vibrational circular
  dichroism: An approach based on the exact factorization of the
  electron-nuclear wave function}.
\newblock {\em The Journal of Chemical Physics}, 143(7):074106, 08 2015,
  arXiv:https://pubs.aip.org/aip/jcp/article-pdf/doi/10.1063/1.4928578/15499312/074106\_1\_online.pdf.

\bibitem{Eich16054110}
F.~G. Eich and Federica Agostini.
\newblock {The adiabatic limit of the exact factorization of the
  electron-nuclear wave function}.
\newblock {\em The Journal of Chemical Physics}, 145:054110, 2016.

\bibitem{Axel163316}
Axel Schild, Federica Agostini, and E.~K.~U. Gross.
\newblock Electronic flux density beyond the born-oppenheimer approximation.
\newblock {\em JOURNAL OF PHYSICAL CHEMISTRY A}, 120(19):3316--3325, MAY 19
  2016.

\bibitem{Scherrer17031035}
Arne Scherrer, Federica Agostini, Daniel Sebastiani, E.~K.~U. Gross, and
  Rodolphe Vuilleumier.
\newblock On the mass of atoms in molecules: Beyond the born-oppenheimer
  approximation.
\newblock {\em Phys. Rev. X}, 7:031035, Aug 2017.

\bibitem{Panati02250405}
Gianluca Panati, Herbert Spohn, and Stefan Teufel.
\newblock Space-adiabatic perturbation theory in quantum dynamics.
\newblock {\em Physical Review Letters}, 88(25):250405, June 2002.

\bibitem{Panati07297}
Gianluca Panati, Herbert Spohn, and Stefan Teufel.
\newblock The time-dependent born-oppenheimer approximation.
\newblock {\em ESAIM: Mathematical Modelling and Numerical Analysis},
  41(2):297--314, March 2007.

\bibitem{Kato50435}
Tosio Kato.
\newblock On the adiabatic theorem of quantum mechanics.
\newblock {\em Journal of the Physical Society of Japan}, 5(6):435--439,
  November 1950.

\bibitem{Tully98407}
John~C. Tully.
\newblock Mixed quantum-classical dynamics.
\newblock {\em Faraday Discussions}, 110:407--419, 1998.

\bibitem{Kapral19998919}
Raymond Kapral and Giovanni Ciccotti.
\newblock Mixed quantum-classical dynamics.
\newblock {\em The Journal of Chemical Physics}, 110(18):8919--8929, may 1999.

\bibitem{Kapral2015073201}
Raymond Kapral.
\newblock Quantum dynamics in open quantum-classical systems.
\newblock {\em Journal of Physics: Condensed Matter}, 27(7):073201, January
  2015.

\bibitem{Diestler20134698}
D.~J. Diestler.
\newblock {Beyond the Born-Oppenheimer Approximation: A Treatment of Electronic
  Flux Density in Electronically Adiabatic Molecular Processes}.
\newblock {\em The Journal of Physical Chemistry A}, 117(22):4698--4708, May
  2013.

\bibitem{Ditler20222448}
Edward Ditler, Tom\'{a}\v{s} Zimmermann, Chandan Kumar, and Sandra Luber.
\newblock {Implementation of Nuclear Velocity Perturbation and Magnetic Field
  Perturbation Theory in CP2K and Their Application to Vibrational Circular
  Dichroism}.
\newblock {\em Journal of Chemical Theory and Computation}, 18(4):2448--2461,
  April 2022.

\bibitem{Abedi10123002}
Ali Abedi, Neepa~T. Maitra, and E.~K.~U. Gross.
\newblock Exact factorization of the time-dependent electron-nuclear wave
  function.
\newblock {\em Phys. Rev. Lett.}, 105:123002, Sep 2010.

\bibitem{Suzuki14040501}
Yasumitsu Suzuki, Ali Abedi, Neepa~T. Maitra, Koichi Yamashita, and E.~K.~U.
  Gross.
\newblock Electronic schr\"odinger equation with nonclassical nuclei.
\newblock {\em Phys. Rev. A}, 89:040501, Apr 2014.

\bibitem{Min15073001}
Seung~Kyu Min, Federica Agostini, and E.~K.~U. Gross.
\newblock Coupled-trajectory quantum-classical approach to electronic
  decoherence in nonadiabatic processes.
\newblock {\em Phys. Rev. Lett.}, 115:073001, Aug 2015.

\bibitem{Li2022113001}
Chen Li, Ryan Requist, and E.~K.~U. Gross.
\newblock Energy, momentum, and angular momentum transfer between electrons and
  nuclei.
\newblock {\em Phys. Rev. Lett.}, 128:113001, Mar 2022.

\bibitem{Arribas2024233201}
Evaristo~Villaseco Arribas and Neepa~T. Maitra.
\newblock Electronic coherences in molecules: The projected nuclear quantum
  momentum as a hidden agent.
\newblock {\em Phys. Rev. Lett.}, 133:233201, Dec 2024.

\bibitem{Abedi1222A530}
Ali Abedi, Neepa~T. Maitra, and E.~K.~U. Gross.
\newblock Correlated electron-nuclear dynamics: Exact factorization of the
  molecular wavefunction.
\newblock {\em The Journal of Chemical Physics}, 137(22):22A530, 2012.

\bibitem{Min2014263004}
Seung~Kyu Min, Ali Abedi, Kwang~S. Kim, and E.~K.~U. Gross.
\newblock Is the molecular berry phase an artifact of the born-oppenheimer
  approximation?
\newblock {\em Phys. Rev. Lett.}, 113:263004, Dec 2014.

\bibitem{Requist16042108}
Ryan Requist, Falk Tandetzky, and E.~K.~U. Gross.
\newblock Molecular geometric phase from the exact electron-nuclear
  factorization.
\newblock {\em Phys. Rev. A}, 93:042108, Apr 2016.

\bibitem{Requist17062503}
Ryan Requist, C\'{e}sar~R. Proetto, and E.~K.~U. Gross.
\newblock Asymptotic analysis of the {Berry} curvature in the {$E\otimes e$}
  {Jahn-Teller} model.
\newblock {\em Physical Review A}, 96(6):062503, December 2017.

\bibitem{Tu2025043075}
Matisse Wei-Yuan Tu and E.~K.~U. Gross.
\newblock Electronic decoherence along a single nuclear trajectory.
\newblock {\em Physical Review Research}, 7(4):043075, October 2025.

\bibitem{FootnoteENAPTapp1}
Ref. \cite{Tu2025043075} focused on electronic decoherence with an involvement
  of the NAPT method without fully elaborating its various aspects. The present
  article is dedicated to a systematic exposition of the method itself.

\bibitem{Longuet-Higgins19581}
Maurice Henry Lecorney~Pryce Hugh Christopher Longuet-Higgins, U.~\"{O}pik and
  R.~A. Sack.
\newblock {Studies of the Jahn-Teller effect .II. The dynamical problem}.
\newblock {\em Proceedings of the Royal Society of London. Series A.
  Mathematical and Physical Sciences}, 244(1236):1--16, February 1958.

\bibitem{Herzberg6377}
G.~Herzberg and H.~C. Longuet-Higgins.
\newblock Intersection of potential energy surfaces in polyatomic molecules.
\newblock {\em Discussions of the Faraday Society}, 35:77, 1963.

\bibitem{OBrien93688}
Mary C.~M. O'Brien and C.~C. Chancey.
\newblock {The Jahn-Teller effect: An introduction and current review}.
\newblock {\em American Journal of Physics}, 61(8):688--697, August 1993.

\bibitem{JoubertDoriol20177365}
Lo\"{i}c Joubert-Doriol and Artur~F. Izmaylov.
\newblock {Molecular "topological insulators": a case study of electron
  transfer in the bis(methylene) adamantyl carbocation}.
\newblock {\em Chemical Communications}, 53(53):7365--7368, 2017.

\bibitem{Ibele202311625}
Lea~M. Ibele, Eduarda Sangiogo~Gil, Basile F.~E. Curchod, and Federica
  Agostini.
\newblock On the nature of geometric and topological phases in the presence of
  conical intersections.
\newblock {\em The Journal of Physical Chemistry Letters}, 14(51):11625--11631,
  December 2023.

\bibitem{Martinazzo2024243002}
Rocco Martinazzo and Irene Burghardt.
\newblock Dynamics of the molecular geometric phase.
\newblock {\em Physical Review Letters}, 132(24):243002, June 2024.

\bibitem{Gossel2019154112}
Graeme~H. Gossel, Lionel Lacombe, and Neepa~T. Maitra.
\newblock On the numerical solution of the exact factorization equations.
\newblock {\em The Journal of Chemical Physics}, 150(15):154112, April 2019.

\bibitem{Gidopoulos-ArX0502433}
Nikitas~I. Gidopoulos and E.~K.~U. Gross.
\newblock Electronic non-adiabatic states.
\newblock 2005, arXiv:cond-mat/0502433.

\bibitem{Gidopoulos201420130059}
Nikitas~I Gidopoulos and E~K~U Gross.
\newblock Electronic non-adiabatic states: towards a density functional theory
  beyond the {Born-Oppenheimer} approximation.
\newblock {\em Philos. Trans. A Math. Phys. Eng. Sci.}, 372(2011):20130059,
  March 2014.

\bibitem{Cohen2025075102}
Galit Cohen, Rachel Steinitz-Eliyahu, E.~K.~U. Gross, Sivan Refaely-Abramson,
  and Ryan Requist.
\newblock Nonadiabaticity from first principles: Exact-factorization approach
  for solids.
\newblock {\em Physical Review B}, 112(7):075102, August 2025.

\end{thebibliography}
\end{document}